\documentstyle[12pt]{article}
\newcommand{\ber}{\begin{eqnarray}}
\newcommand{\eer}{\end{eqnarray}}
\newcommand{\bea}{\begin{equation}}
\newcommand{\eea}{\end{equation}}
\begin{document}
\title{Statistical Approach to Gene Evolution} 
\author{$Sujay\:Chattopadhyay^{*},\:William\:A.\:Kanner\:and\:Jayprokas\:Chakrabarti$\\\\
 Department of Theoretical Physics, Indian Association for the Cultivation of \\
Science, Calcutta 700 032, INDIA.}
\date{}
\maketitle
\begin{abstract}
The evolution in coding DNA sequences brings new flexibility and freedom to the codon words, even as the underlying nucleotides get significantly ordered. These curious contra-rules of gene organisation are observed from the distribution of words and the second moments of the nucleotide letters. These statistical data give us the physics behind the classification of bacteria.
\end{abstract}
\newpage
Over the years the statistical approach to genes has become prominent. The hidden Markov models are used in the alignment routines of biological sequences [1]. For the secondary structures of the sequences stochastic context-free and context-sensitive grammars [2] are applied [3]. The recent discovery of the fractal inverse power-law correlations [4] in these biological chains have led to ideas that statistically these sequences have features of music and languages [5-7]. Languages evolve with time. The vocabulary increases; the rules that dominate get progressively optimised so the order and information content is more. The purpose of this work is to track the statistical basis of the evolution in the coding DNA sequences (CDS).  
\par
The CDS are multiple of 3-tuples, the codons. The nucleotides adenine (A), cytosine (C), guanine (G) and thymine (T) taken in groups of three work to build the amino acid chains called proteins. The word-structure of CDS is, therefore, well known. We want to study evolution in terms of these words, their distributions and the moments.  
\par
It is known that any prose does not carry all the ingredients of evolution of languages. Similarly the CDS of any gene does not have all the salient features that accompany change. The genes that are present in the whole range of organisms, from the lowest bacteria to the highest mammals, and therefore connected to fundamental life processes are normally considered to be best suited to function as evolutionary markers. With this in view we choose glyceraldehyde-3-phosphate dehydrogenase (GAPDH) CDS for its ubiquitous presence in all living beings. The enzyme it codes for catalyses one of the crucial energy-producing steps of glycolysis, the common pathway for both aerobic and anaerobic respiration.  
\par
Distribution of words is studied for languages. The frequency of words is plotted against the rank. Here the total number of occurrences of a particular word is termed its frequency. The word most frequent has rank=1, the next most has rank=2, and so on. For natural languages, the plot gives the Zipf [8] behaviour:
\bea
f_{N}\:=\:\frac{f_{1}}{N}
\eea
where $N$ stands for the rank and $f_1$ and $f_N$ are the frequencies of words of rank 1 and $N$ respectively. The Zipf-type approach to the study of DNA has brought methods of statistical linguistics into DNA analysis [6]. The generalized Zipf distribution of n-tuples has provided hints that the DNA sequences may have some structural features common to languages. In this work we confine ourselves to the distribution of 3-tuples, the codons, in the CDS. The words, therefore, are non-overlapping and on the single reading frame.    
\par  
The frequency-vs-rank plot of the codon words show that these distributions, given the frequency of rank 1 and the length of the sequence, are almost completely defined through the universal exponential functional form [9]:
\bea
f_{n}\:=\:f_{1}.e^{-\beta(N-1)}
\eea
The parameter, called $\beta$, is determined by the ratio
\bea
\beta\:\approx\:\frac{f_{1}}{L}
\eea
$\beta$ measures the frequency of rank 1 per unit length of the sequence. The exponential form (2) is to be compared to the usual Boltzmann distribution. The rank of the word is akin to energy; $\beta$ is analogous to inverse temperature. The relationship (3) that $\beta$ is frequency of rank 1 per unit length is supported well from data [9]. The analogy between word distributions and the classical Boltzmann concepts goes deeper. A decrease in $\beta$, from (3), implies frequency of rank 1 per unit length goes down. In that case the vocabulary clearly increases. More words are used, thereby more states are accessed. For the GAPDH CDS we find the evolution is driving it to higher temperatures; into more freedom for words, into more randomisation. $\beta$ evolves monotonically.
\par
Underneath, however, there runs a curious counterflow. Suppose we look into the nucleotides that constitute the sequence, once again in windows of size 3 and in the same reading frame. First, we ask how much order there is in the sequence. To find out we study the second moments of the letters A, C, G and T. These second moments, by themselves, do not produce any pattern. The GAPDH CDS has about 1000 bases. For each organism the proportions of A, C, G and T in the GAPDH CDS are different. This strand-bias, interestingly, masks a remarkable underlying trend.
\par
To get there the strand-bias has to be eliminated. The order in the sequence, we assume, is its deviation from the random. We define the quantity $X$, a measure of this deviation, as follows:\\
\begin{center}
$X$ = {$\frac{Second\:Moment\:of\:the\:Base\:Distribution\:in\:GAPDH\:CDS}{Second\:Moment\:of\:the\:Base\:Distribution\:in\:the\:random\:sequence\:with\:identical\:strand\:bias}$}
\end{center}
Normalised as above, the effect of the strand bias is unmasked. $X$ values of GAPDH change monotonically with evolution. The data tells us there is an increase in persistence amongst the letters (in windows of size 3) with evolution in the CDS [10].
\par
The evolution in the GAPDH CDS is then the result of these two contra trends: while words acquire greater uniformisation, the underlying letters have more order. The monotonic behaviours of $\beta$ and $X$ with evolution give us the physics behind the biological classification of bacteria.      
\\\\
$\bf{Methods}$\\
$\underline{Word\:\:Distributions}$
\\
For the codons it is known [9] the exponentials give somewhat better fits over the usual power laws. The exponential form, equation (2), is characterized by the parameter $\beta$. The quantity has some universal features in that it is almost completely determined by $f_1$ and the length of the CDS. The relationship [9]
\bea
\beta\:=\:\frac{f_1-1}{L}\:+\:\frac{1}{2}.\frac{(f_1-1)^2}{L^2}
\eea
is known to fit observations on diverse genes. For the bacterial GAPDH CDS the results of $\beta$ are given in Table 2.
\\
$\underline{Moments}$
\\
Consider the 4-dimensional walk model [11,12] such that A, C, G and T correspond to unit steps, in the positive direction, along $X_A$, $X_C$, $X_G$ and $X_T$ axes. After $n$-steps if the co-ordinate of the walker is
($n_A$, $n_C$, $n_G$, $n_T$), then, clearly,
\bea
n\:=\:n_A+n_C+n_G+n_T
\eea
and $n_i$ ($i$ $\equiv$ A,C,G,T), is the number of nucleotide of type $i$ in the sequence just walked.
\par
If the sequence has $n$ bases, and $n_i$ is the number of base of type $i$, the strand bias of the sequence is the proportion of $n_i$ in $n$, defined as
\bea
p_i\:=\:\frac{n_i}{n}
\eea
The probability distribution for the single step in this 4-d walk is
\bea
P_1(x)\:=\:\sum_{i}p_{i}\delta(x_i-1)
\eea
where $\delta$ is the usual $\delta$-function of Dirac. The characteristic function of the step is the Fourier transform of equation (7),
\bea
P_1(k)\:=\:\sum_{i}p_{i}e^{ik_i}
\eea
The characteristic function of $l$ steps
\bea
P_l(k)\:=\:[P_1(k)]^l
\eea
The second moments (i.e. the average values) of distributions may be obtained taking derivatives of $P_l(k)$ with respect to $k$. Thus for the random sequence (indicated by the subscript $r$) with the strand bias (6), we get the average values: 
\bea
<n^2_i>_r\:=\:l[(l-1)p^2_i+p_i]
\eea
\bea
<n_in_j>_r\:=\:l(l-1)(p_i.p_j)\:\:\:\:\:\:\:\:\:\:\:\:\:\:\:(i \neq j)  
\eea
\par
We are interested in codons, therefore, the window size $l$ in equations (10) and (11) is chosen to be 3. For the actual sequences we calculate $<n^2_i>_{seq}$ and $<n_in_j>_{seq}$. The quantities
\bea
X_D\:=\:\frac{<n^2_i>_{seq}}{<n^2_i>_r} 
\eea 
[where D $\equiv$ AA,CC,GG,TT]\\ 
and
\bea
X_{OD}\:=\:\frac{<n_in_j>_{seq}}{<n_in_j>_r}\:\:\:\:\:\:\:\:\:\:\:\:\:\:\:(i \neq j)  
\eea
[where OD $\equiv$ AC,AG,AT,CG,CT,GT]\\
measure the deviation of the diagonal and off-diagonal second moments of the sequence to those of the random sequence of identical strand bias respectively. Finally, we come up with an over-all averaged index, $X$, given by
\bea
X\:=\:\frac{\sum{X_D}\:+\:\sum{X_{OD}}}{10}
\eea
This $X$ provides a measure of the order in the sequences. 
\newpage
\noindent
$\bf{Observations\:and\:Results}$\\
To set the basis for what we discuss later, we begin by recording the $\beta$ and the $X$ values of higher organisms, the eukaryotes (Table 1).   We confine our discussion of the eukaryotes to three broad categories: fungi, invertebrates and vertebrates. It is known [13] from fossil records the oldest fungi came about 900 million years (Myr) before present (bp). The oldest fungal species, identified with certainty, are from the Ordovician period, i.e., some 500 Myr bp. The fossil records of invertebrates suggest this group came about the same time as the fungi. The vertebrates came later, about 400 Myr bp, in late Ordovician and Silurian period.
\par
Let us look at the $\beta$ and the $X$ values of these eukaryotic groups. Fungi has the highest, followed by invertebrates, while for the vertebrates the $\beta$ and the $X$ reach minima. We conclude the $\beta$ and the $X$ decrease with evolution. The data further suggest fungi and invertebrates came about the same time and underwent parallel evolution, while the representives of the vertebrate group came later in the evolutionary line-up.
\par
Having set the basis, let us now look at 14 bacterial species from three groups: cyanobacteria, proteobacteria (that includes vast majority of gram-negative bacteria), and the ${\it{Bacillus/Clostridium}}$ group, a type of gram-positive bacteria. Table 2 summarises the $\beta$ and the $X$ values of these samples. These bacterial groups arose during the Precambrian period of geological time-scale, but there are several schools of thought regarding their specific times of origin within this period.
\par
We approach the bacterial GAPDH CDS with two differing statistical measures, the $\beta$ and the $X$. Interestingly, both give us almost identical trends (Figs. 1 and 2). ${\it{Lactobacillus\:delbrueckii}}$, a member of the ${\it{Bacillus/Clostridium}}$ group, has the highest $\beta$ and $X$ values (Table 2). There is then a large measure of overlap between the ${\it{Bacillus/Clostridium}}$ group and the proteobacteria (Figs. 1 and 2). The extent of overlap of the $\beta$ values is somewhat more than that of the $X$. The cyanobacterial samples have the minimum values of the $\beta$ and the $X$. There is no overlap between the cyanobacterial values of the $\beta$ and the $X$ with the ${\it{Bacillus/Clostridium}}$ group. The overlap between the proteobacteria and the cyanobacteria is small. Only one proteobacterial sample, ${\it{Brucella\:abortus}}$ has greater $\beta$ value than the cyanobacterial member, ${\it{Synechocystis}}$ sp. (strain PCC 6803).
\par
The averages of the $\beta$ or the $X$ has the maximum value in the ${\it{Bacillus/Clostridium}}$ group, followed by the proteobacteria, while the cyanobacteria samples have the lowest values. In line with our observations on the eukaryotes, we propose (Figs. 1 and 2) that the ${\it{Bacillus/Clostridium}}$ group originated some time before the proteobacterial species, but later both groups evolved in parallel. The cyanobacterial samples are of recent origin compared to these groups. The trends in the $\beta$ and the $X$ give us identical patterns that segregate the bacterial species into groups. Amusingly, the results seem to be in agreement with what is accepted so far regarding the phylogenetic relationships among these three groups [14]. Our study of the GAPDH CDS, its word distributions, and the moments gives us the physics underlying evolution.
\par
The decrease in $\beta$ with evolution for the GAPDH CDS tells us that evolution is taking the gene progressively towards higher temperatures. The $\beta$ value, we recall, is the frequency of rank one per unit length. Lowering of the $\beta$ implies less dominance of the maximum weight. In consequence, the other words enjoy greater freedom, the vocabulary increases and more states are accessed. In a sense the evolution in the GAPDH CDS mirrors Boltzmannian statistics. Even though the GAPDH CDS has evolved in a complex evolutionary regime in contact with environment, the Boltzmannian behaviour is useful. For instance, it allows us to define the word-entropy of the CDS. That gives us a measure of the information content of the words in biological chain. 
\par
At the level of the nucleotide letters A, C, G and T, the order is measured by the quantity $X$. As we look into the diagonal averages $X_{D}$, (12), we find it increases with evolution. For the window of size 3, this growing diagonal moment implies a rising persistent correlation. In consequence, the off-diagonal averages $X_{OD}$, (13), go down, decreasing antipersistence. Looked at from the letters, the sequences become less uniform and deviate more from the random sequence of identical strand bias. The order, or the information, in the arrangement of letters shows a rising trend with evolution.  
\par
Does any CDS that is an evolutionary marker evolve in ways similar to the GAPDH? We have worked with the CDS of some other glycotic enzymes, such as phosphoglycerate kinase, and found they behave similarly. Other evolutionary markers such as the ribulose-1,5-bisphosphate carboxylase/oxygenase enzyme large segment (rbcL) show similar behaviour. We use these data for biological subclassification. The CDS for ribosomal RNA is another class of sequence that is being investigated. It does not code for protein, but for RNA, and has periods other than 3. The 3 period does exist, but is not predominant.
\par
Sequence modeling has recently become important. The fractal correlations in the sequences led to the expansion-modification system [15]. Later came the insertion models [16]. Here the differences in the CDS and non-coding sequences were observed and the non-coding sequences modeled. The unifying models of copying-mistake-maps [17] modeled both the coding and the non-coding parts. In these models the statistical features of the non-coding sequences have received emphasis. The evolutionary features of the GAPDH CDS isolates the statistical aspects that underlie evolution in coding sequences. The statistics of the word distributions and the subtle cross current of the second moments, we hope, will lead further in these efforts.
\\\\
$ {\bf{Acknowledgments}} $ \\
S.C. thanks Professor Anjali Mookerjee for many discussions. W.A.K. is supported by the John Fulbright foundation in the laboratory of J.C.               
\newpage
$^*$Electronic address: tpsc@mahendra.iacs.res.in

\newpage\noindent
$ {\bf{Figure\:Legends}} $ \\\\
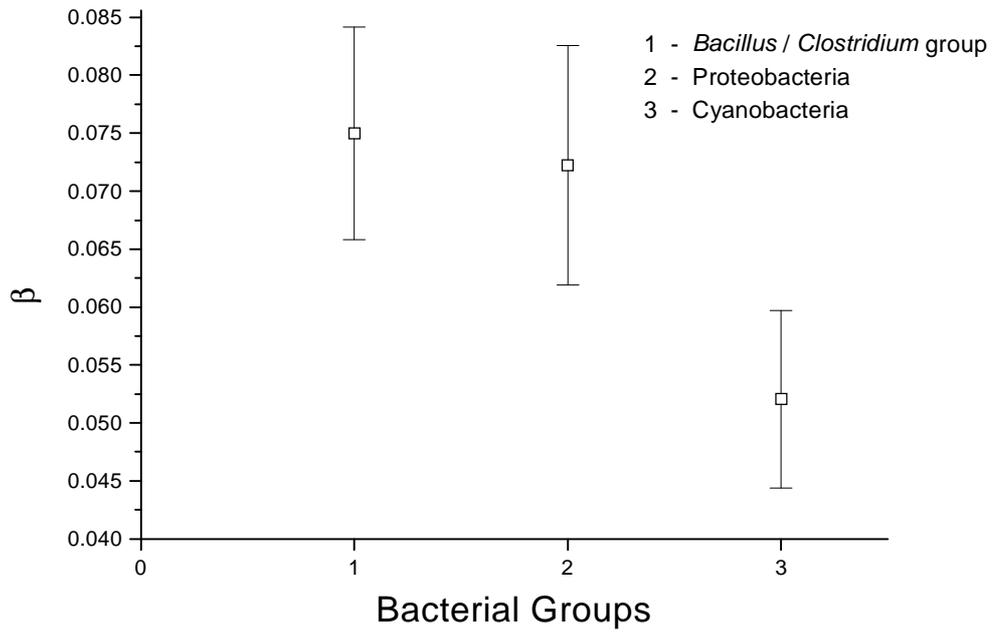
Figure 1. The average $\beta$ values for the GAPDH CDS from three bacterial groups (see Table 3). The error bars indicate the standard deviation from the average values. 
\\\\
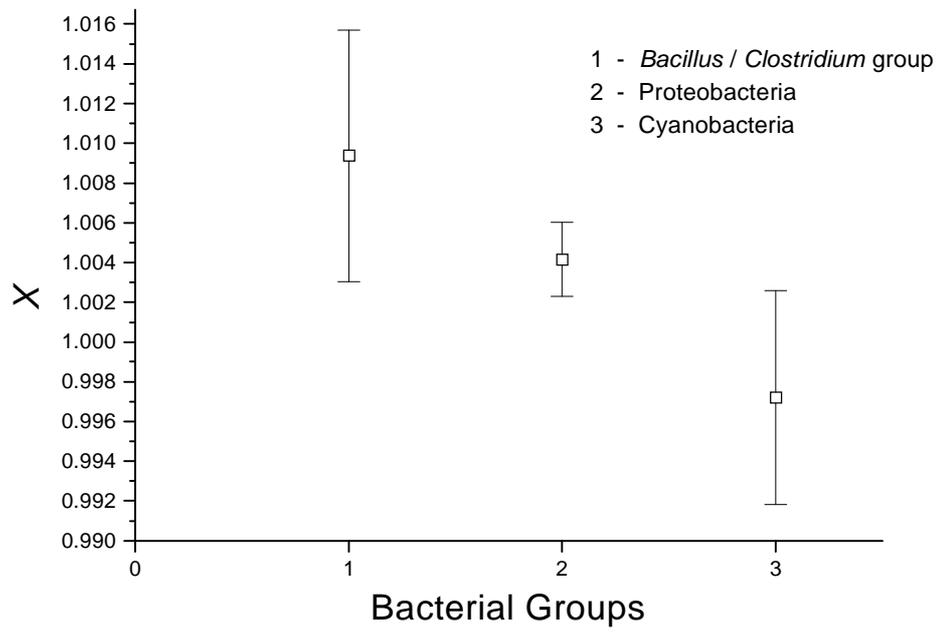
Figure 2. The average $X$ values for the GAPDH CDS from three bacterial groups (see Table 3). The error bars indicate the standard deviation from the average values.

\newpage
\begin{table}
\renewcommand{\arraystretch}{1.5}
\begin{center}
\caption{\bf{The average $\beta$ and $X$ values of GAPDH CDS for eukaryotic groups, along with the range of deviations in the respective groups.}} 
\bigskip
\begin{tabular}{|c c c|}
\hline
Group & $\beta$ & $X$  \\
\hline
Vertebrates & 0.05398 ($\pm$0.00414) & 0.99698 ($\pm$0.004) \\
Invertebrates & 0.07503 ($\pm$0.01067) & 1.00235 ($\pm$0.00261) \\
Fungi & 0.07742 ($\pm$0.00389) & 1.00705 ($\pm$0.00175) \\
\hline
\end{tabular}
\end{center}
\renewcommand{\arraystretch}{1.5}
\end{table}

\newpage
\begin{table}
\renewcommand{\arraystretch}{1.5}
\begin{center}
\caption{\bf{The $\beta$ and the $X$ values of the GAPDH CDS from the bacterial species that have been used in our study (source: GenBank and EMBL databases).}} 
\bigskip
\begin{tabular}{|c c c c c|}
\hline
Organism & Accession No. & Group & $\beta$ & $X$ \\
\hline
${\it{Bacillus\:megaterium}}$ & M87647 & ${\it{Bacillus/Clostridium}}$ & 0.07662 & 1.01185  \\
${\it{Bacillus\:subtilis}}$ & X13011 & ${\it{Bacillus/Clostridium}}$ & 0.07431 & 1.00912  \\
${\it{Clostridium\:pasteurianum}}$ & X72219 & ${\it{Bacillus/Clostridium}}$ & 0.07837 & 1.00483  \\
${\it{Lactobacillus\:delbrueckii}}$ & AJ000339 & ${\it{Bacillus/Clostridium}}$ & 0.08529 & 1.01861 \\ ${\it{Lactococcus\:lactis}}$ & L36907 & ${\it{Bacillus/Clostridium}}$ & 0.06038 & 1.00245  \\
${\it{Pseudomonas\:aeruginosa}}$ & M74256 & Proteobacteria & 0.08166 & 1.00338  \\
${\it{Escherichia\:coli}}$ & X02662 & Proteobacteria & 0.08366 & 1.00447  \\
${\it{Brucella\:abortus}}$ & AF095338 & Proteobacteria & 0.05713 & 1.00604  \\
${\it{Zymomonas\:mobilis}}$ & M18802 & Proteobacteria & 0.07721 & 1.00457  \\
\hline
\end{tabular}
\end{center}
\renewcommand{\arraystretch}{1.5}
\end{table}

\newpage
\begin{table}
\renewcommand{\arraystretch}{1.5}
\begin{center}
\begin{tabular}{|c c c c c|}
\hline
Organism & Accession No. & Group & $\beta$ & $X$ \\
\hline
${\it{Rhodobacter\:sphaeroides}}$ & M68914 & Proteobacteria & 0.06539 & 1.00564  \\
${\it{Xanthobacter\:flavus}}$ & U33064 & Proteobacteria & 0.06839 & 1.00086  \\
${\it{Anabaena\:variabilis}}$ & L07498 & Cyanobacteria & 0.04547 & 1.00073  \\
${\it{Synechococcus}}$ PCC 7942 & X91236 & Cyanobacteria & 0.05025 & 0.99988  \\
${\it{Synechocystis}}$ PCC 6803& X83564 & Cyanobacteria & 0.06043 & 0.99101  \\
\hline
\end{tabular}
\end{center}
\renewcommand{\arraystretch}{1.5}
\end{table}

\newpage
\begin{table}
\renewcommand{\arraystretch}{1.5}
\begin{center}
\caption{\bf{The average $\beta$ and $X$ values of the GAPDH CDS for the three bacterial groups, along with the range of deviations in the respective groups.}} 
\bigskip
\begin{tabular}{|c c c|}
\hline
Group & $\beta$ & $X$  \\
\hline
${\it{Bacillus/Clostridium}}$ & 0.07499 ($\pm$0.00914) & 1.00937 ($\pm$0.00633) \\
Proteobacteria & 0.07224 ($\pm$0.01033) & 1.00416 ($\pm$0.00187) \\
Cyanobacteria & 0.05205 ($\pm$0.00764) & 0.99721 ($\pm$0.00538)  \\
\hline
\end{tabular}
\end{center}
\renewcommand{\arraystretch}{1.5}
\end{table}

\end{document}